\documentclass[traditabstract]{aa}

\usepackage{txfonts}

\usepackage{graphicx}

\usepackage[cmyk]{xcolor}

\usepackage{ulem}

\usepackage{natbib}
\bibpunct{(}{)}{;}{a}{}{,} 

\begin{document}

\title{Crossing barriers in planetesimal formation: The growth of mm-dust aggregates with large constituent grains}

\author{Tim Jankowski\inst{1}\thanks{tim.jankowski@uni-due.de} \and Gerhard Wurm\inst{1} 
\and Thorben Kelling\inst{1} \and Jens Teiser\inst{1} \and Walter Sabolo\inst{2}, Pedro J. Guti{\'e}rrez\inst{2} \and Ivano Bertini\inst{3}}

\institute{Fakult\"at f\"ur Physik, Universit\"at Duisburg-Essen, Lotharstrasse 1, 47057 Duisburg, Germany \and Instituto de Astrof\'isica de Andaluc\'ia, CSIC, Glorieta de la Astronomia s/n, 18008 Granada, Spain \and Centro Interdipartimentale di Studi e Attivit\'a Spaziali (CISAS) ``G. Colombo'', Universit\'a di Padova, Via Venezia 15, 35131 Padova, Italy}

\abstract{Collisions of mm-size dust aggregates play a crucial role in the early phases of planet formation. It is for example currently unclear whether there is a bouncing barrier where millimeter aggregates no longer grow by sticking. We developed a laboratory setup that allowed us to observe collisions of dust aggregates levitating at mbar pressures and elevated temperatures of 800~K. 
We report on collisions between basalt dust aggregates of from 0.3 to 5~mm in size at velocities between 0.1 
and 15~cm/s. Individual grains are smaller than 25~$\mu$m in size. 
We find that for all impact energies in the studied range sticking occurs 
at a probability of $32.1\pm 2.5$~\% on average. In general, the sticking probability decreases with increasing impact 
parameter. The sticking probability increases with energy density (impact
energy per contact area). We also observe collisions of aggregates 
that were formed by a previous sticking of two larger aggregates. 
Partners of these aggregates can be detached by a second collision with a probability of on average $19.8\pm 4.0$~\%.
The measured accretion efficiencies are remarkably high compared to other experimental results. We attribute this to the relatively large dust grains  used in our experiments, which make aggregates more susceptible to restructuring and energy dissipation. Collisional hardening by compaction might not occur as the aggregates are already very compact with only $54$~\% $ \pm 1$~\% porosity. The disassembly of previously grown aggregates in collisions might stall further aggregate growth. However, owing to the levitation technique and the limited data statistics, no conclusive statement about this aspect can yet be given. We find that the detachment efficiency decreases with increasing velocities and accretion dominates in the higher
velocity range. For high accretion efficiencies, our experiments suggest that continued growth in the mm-range with larger constituent grains would be a viable way to produce larger aggregates, which might in turn form the seeds to proceed to growing planetesimals.}


\titlerunning{The growth of mm-dust aggregates with large constituent grains}

\maketitle

\section{Introduction}
Terrestrial planet formation is assumed to begin with the aggregation of micrometer-sized dust particles in protoplanetary disks. This initial growth is 
governed by cohesive forces \citep{blum2008}. The relative velocities
between particles are determined by the different types of gas-grain coupling \citep{weidenschilling1977}. 
At low collision velocities on the order of a few mm/s, fractal dust aggregates form \citep{blum1996, dominik1997}. This first phase can be characterized as a hit-and-stick regime, as all collisions lead to rigid sticking at the first contact point \citep{wurm1998, bertini2009}. Eventually, energies become high enough for restructuring
and porous but non-fractal aggregates form \citep{blum2000}. Collision velocities then increase with aggregate size \citep{weidenschilling1993}. At a certain stage, mm-sized compact dust agglomerates of porosities on the order of 60 -- 70~\% might have formed \citep{weidling2009}.  It is currently unclear how planet formation proceeds from here. 
Depending on the disk model, the relative velocities rapidly increase from $10^{-3}$ m/s or $10^{-2}$ m/s to several tens of m/s at dm to m size \citep{desch2007, weidenschilling1993}.
Numerical models cover the high speed parts and, depending on the assumed parameters such as porosity, elastic properties or collision velocities, further growth is or is not possible \citep{geretshauser2011, schaefer2007, wada2009}. 
However, it appears non-trivial to reach the high speed collisions that partly permit growth. Recent experiments have shown that the collision results for mm-sized dust aggregates are more complex and growth can occur \citep{weidling2011}. \citet{weidling2011} report a sticking probability of {5.6~\%} at collision velocities of between 3 cm/s and 10 cm/s that however decreases with velocity. Experimental studies in general have so far found that bouncing is a typical outcome after the sticking phase \citep{guettler2010}. With the assumption that colliding particles  bounce off each other, further growth would be prevented. This assumption was introduced by \citet{zsom2010} as the bouncing barrier. The underlying physics is that these collisions are more or less elastic and not enough energy can be dissipated by the compact aggregates to allow sticking. This is also the reason why growth is again possible at higher collision velocities as fragmentation leads to enough dissipation to allow mass transfer and net growth \citep{wurm2005}.

One promising way to overcome the problems of collisional growth is the trapping of solid particles by turbulence. If the local particle density {in the protoplanetary disk} is increased sufficiently, gravitational instabilities lead to the rapid formation of planetesimals \citep{johansen2007}. Particles might also be concentrated by a streaming instability \citep{youdin2007}. Current models describing this growth mechanism are based on the assumption that most of the solid material in protoplanetary disks is trapped in macroscopic (decimeter size) dust agglomerates. This size range therefore has to be reached by aggregation. If bouncing prevents growth, then these instability scenarios are also unable to solve the problem of planet formation.  

In experimental studies, it has been shown that the collision characteristics change significantly, when bodies of different size collide with each other. \citet{wurm2005}, \citet{teiser2009b}, and \citet{teiser2011a} demonstrated that the fragmentation of the smaller body can lead to a partial mass transfer from the smaller to the larger body if the impact velocity is high enough for projectile fragmentation. Once the aggregates have reached a size of a few centimeters, the reaccretion of dust by gas drag can efficiently enhance the growth rate \citep{wurm2001, teiser2009a, teiser2011b}. Mass transfer by (projectile) fragmentation as well as growth by reaccretion require a certain amount of aggregates to be already in the centimeter range. To overcome the bouncing barrier, it remains uncertain how at least a few particles might cross the bouncing barrier to form cm-size aggregates.  

In \citet{weidling2011}, the monomer size was between 0.5\,$\mu$m and 10\,$\mu$m (with 80 \% between 1 $\mu$m and 5 $\mu$m). The experimental studies of \citet{blum2006} showed that the mechanical properties of dust aggregates are determined by the size distribution of the monomers. Depending on the monomer size, the restructuring of dust aggregates during collisions will take place at varying collision velocities. Experimental evidence that the size distribution of the monomers influences the outcome of aggregate collisions was also presented by \citet{langkowski2008}. Although the focus of this study was on collisions between highly porous aggregates of different sizes (mm vs. cm), it was clearly shown that thresholds for sticking and/or fragmentation depend on the size distribution. 

\begin{figure}
\resizebox{\hsize}{!}{\includegraphics{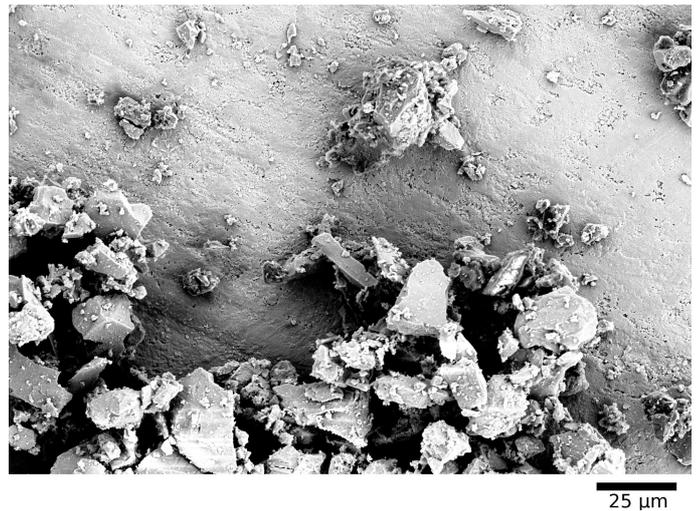}}
\caption{Scanning electron microscope image of the basalt sample used. The mass is dominated by monomers of size $\gtrsim 20 \mu$m.}
\label{fig:basaltem}
\end{figure}

Here, we present an experimental study of the collisions between dust aggregates with sizes between 0.3 and 5~mm at collision velocities of between 0.1 cm/s and 15 cm/s. This parameter range is comparable to the experiments of \citet{weidling2011}, equaling that of the bouncing barrier proposed by \citet{zsom2010}. In comparison to previous aggregation studies \citep{blum2008, guettler2010, weidling2011}, the monomer size is significantly larger. Within this study, we use basalt with a broad size distribution of between 0.1 $\mu$m and 25 $\mu$m. As the basalt consists mostly of silicates, we regard this material as a suitable analogue for protoplanetary dust. In Fig. \ref{fig:basaltem}, we present a scanning electron microscope (SEM) image of the  dust. The aforementioned size range for basalt contains the largest of the grains regularly found in meteorites \citep{brearley1999}. Some authors argue that these are of nebular origin, though others also note that these large grains form on asteroids \citep{scott2005}. Particles found on asteroid Itokawa are within the size range \citep{nakamura2011} and large particles were also found on comet Wild 2 \citep{rietmeijer2008}. With the exception of processed particles such as chondrules, tens of micrometer are at the larger end of the particle size range.
However, this extreme might hold a key to providing the seeds for the growth of larger aggregates. 

\section{Experimental setup}
\begin{figure}
\resizebox{\hsize}{!}{\includegraphics{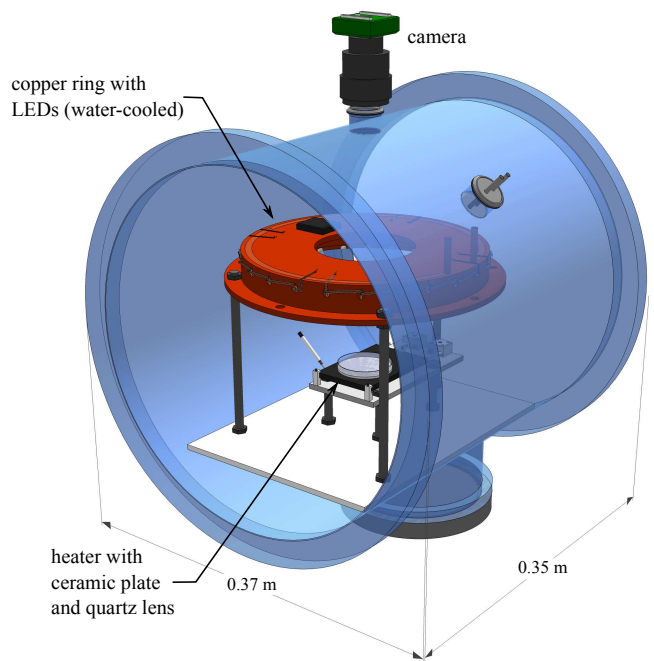}}
\caption{Experimental setup. A heater is placed within a vacuum chamber. For contrast enhancement a ceramic plate and for confinement a concave quartz lens are placed on the heater. Dust aggregates levitate above the lens and are observed by a camera.}
\label{fig:setup1}
\end{figure}

The experimental setup is shown in Fig. \ref{fig:setup1}. Within a vacuum chamber, a heater (up to 800 K), and an illumination source (LED ring) are installed. A camera observes the collisions from above. To enhance the contrast, a white ceramic plate is placed on the heater. To confine
the colliding dust aggregates to the field of view, a slightly concave quartz glass lens is placed on top of the ceramic plate.

\begin{figure}
\resizebox{\hsize}{!}{\includegraphics{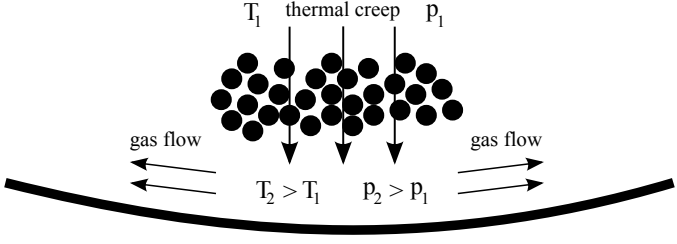}}
\caption{Principle of the Knudsen levitation. The pores of the aggregate act as a collection of micro-channels. Thermal creep is induced by the temperature difference over the aggregate. The pressure increases below the aggregate, and the aggregate is then levitated. The levitation height is limited by the gas flow to the sides, which reduces the overpressure once the aggregate has been levitated \citep{kelling2009}.}
\label{fig:knudsen}
\end{figure}

\begin{figure}
\resizebox{\hsize}{!}{\includegraphics{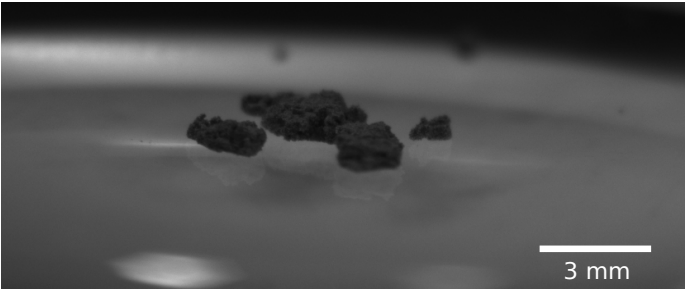}}
\caption{Example of levitating dust aggregates (19 mbar and 798 K heater temperature).}
\label{fig:dust}
\end{figure}

Basalt aggregates of millimeter size, composed of particles $<25$ $\mu$m, are placed loosely onto the glass. In general, fewer than 15 aggregates are placed simultaneously on the lens. The vacuum chamber is then evacuated. Experiments were carried out at $\sim$ 20 mbar and $\sim$ 800 K. In these conditions, the dust aggregates levitate. This mechanism was discovered only recently \citep{kelling2009}. While gravity is still present, there is no interaction (friction, sticking) with any surface and the interaction between different aggregates can be observed.
Levitation is induced by a pressure increase due to thermal creep through the pores of the aggregates.  
Fig. \ref{fig:knudsen} illustrates the principle of the Knudsen levitation mechanism and Fig. \ref{fig:dust} shows a side view of levitating aggregates.

\citet{knudsen09} found that if two gas reservoirs at different temperatures are connected by a capillary tube whose diameter $d_{\rm t}$ is small compared to the mean free path $\lambda$ of the gas molecules ($d_{\rm t}\ll \lambda$), the pressure increases at the warmer side. In equilibrium, the
pressure ratio is given by $p_2/p_1=\sqrt{T_2/T_1}$, where $p_2$ and $T_2$ are the pressure and temperature in the warmer chamber and $p_1$ and $T_1$ are the pressure and temperature in the colder chamber. \citet{muntz2002} showed that at $Kn = \lambda \cdot d_{\rm t}^{-1} \simeq 1$ the pressure difference $\Delta p = \left| p_1 - p_2 \right|$ is
\begin{equation}
\Delta p = p_{\rm avg} \frac{Q_{\rm T}}{Q_{\rm P}}\frac{\Delta T}{T_{\rm avg}},
\label{eq:muntz}
\end{equation}
where $p_{\rm avg}$ and $T_{\rm avg}=(T_1+T_2)/2$ are the average pressure and temperature, $Q_{\rm T}/Q_{\rm P}$ is the ratio of the coefficients of the thermal creep to back flow of the gas and $\Delta T = \left| T_1 -T_2\right|$ is the temperature difference.

The lateral extensions of the dust samples are much smaller than the glass cavity. Hence, the bottom temperature of the dust aggregates is determined by the heater temperature $T_2$. At the top, the aggregates cool by thermal radiation. The loss of heat is counterbalanced by thermal conduction through the aggregates. In equilibrium, the temperature $T_1$ at the top is given by 
\begin{equation}
\sigma T_1^4 = \kappa_{\rm p} \frac{T_2-T_1}{l},
\label{eq:toptemp} 
\end{equation}
where $\sigma = 5.67 \times 10^{-8}$ W m$^{-2}$ K$^{-4}$ is the Stefan-Boltzmann constant, $\kappa_{\rm p}$ is the aggregate's thermal conductivity, and $l$ is the aggregate's vertical extension. The dust aggregates have pores, hence the aggregates act as a collection of micro-channels. At the pressures used in the experiments, the mean free path of gas molecules is comparable to the mean pore size of the aggregates.

The micro-channels formed by the pores of an aggregate, have different temperatures at their inlet (top of the aggregate) and outlet (bottom of the aggregate). According to Eq.(\ref{eq:muntz}), thermal creep leads to a pressure increase $\Delta p$ below the aggregate. If the force $F_{\rm kn}=\Delta p\,A$ acting at the bottom side $A$ of the aggregate compensates gravity $F_{\rm G}$, the aggregate is lifted. With rising height, gas is released to the open sides below the aggregate. The levitation height is limited to some tens to a hundred of micrometers as then no additional overpressure can be established because the gas below the aggregate can escape to the sides. As the aggregates used in this experiment have typical vertical extensions $\gg 100 \,\mu$m and the levitation heights are more or less equal, the setup can be treated as two-dimensional. 
The strength of the levitation can be estimated as follows. We consider an aggregate with an area of \mbox{$A =$ 4~mm$^2$}, a vertical extension of \mbox{$l =$ 1~mm}, and a density of {1332}~kg~m$^{-3}$, which results in an aggregate mass of 5.33 mg. The pressure increase below the aggregate can be calculated in an analogous way to \citet{kelling2009}. Using spherical, 20~$\mu$m sized particles arranged in a simple cubic lattice, the resulting micro-channels have a minimum diameter of $d_{\rm t} = $ 8.3~$\mu$m. With the mean free path of the gas molecules of $\lambda = 9.32 \, \mu$m at a 780~K heater temperature and 19~ mbar pressure, the ratio of the coefficients of the thermal creep to back flow of the gas is $Q_{\rm T}/Q_{\rm P}= 0.22$ \citep{sone1990}. The top temperature of the aggregate is given by Eq.(\ref{eq:toptemp}), resulting in $T_1 = 667$~K. With $T_{\rm avg}  = (T_1 + T_2) / 2 = 724$~K and $p_{\rm avg} \approx p_1 = 19$~mbar, the induced overpressure below the aggregate (Eq.(\ref{eq:muntz})) is $\Delta p = 0.65$ mbar. The force acting on the aggregate from below is {$F_{\rm kn} = \Delta p ~ A = 2.6 \times 10^{-4}$~N}. With $F_{\rm G} = m\,g$ and $g=9.81$~m/s$^2$, the ratio of the lifting force to gravity is
\begin{equation}
\frac{F_{\rm kn}}{F_{\rm G}} = 5.
\label{eq:ratio}
\end{equation}

\section{Experiments}
To characterize the single aggregate collisions, a set of parameters of every collision is determined: the masses of the colliding aggregates $m_1$ and $m_2$, the aggregate shapes (described by the circularity $c$), the collision velocity {$v$}, the maximum contact area $A_{\rm c}$ between the colliding aggregates, the impact parameter {$I$}, and resulting parameters such as either the impact energy $E$ or the impact energy density $E_{\rm d}$ (defined below).

{\textit{Aggregate masses} -- The projected aggregate areas were measured directly from the two-dimensional (2D) video recordings. Based on these measurements we model the aggregate volume by assuming a spherical hemisphere (steep rise to the edge in Fig. \ref{fig:volumefit}). We compared the resulting vertical structure of the particles with side view images from the aggregates. The model reproduces the shape and hence the mass to within an accuracy of about $50$ \%.

\begin{figure}
\resizebox{\hsize}{!}{\includegraphics{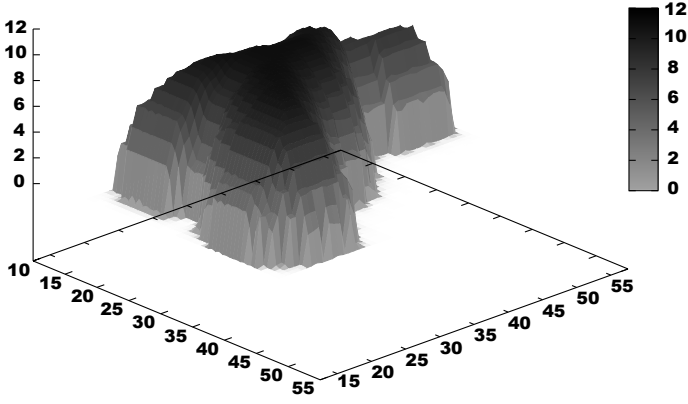}}
\caption{Example of an aggregate volume approximated by a hemisphere (arbitrary units).}
\label{fig:volumefit}
\end{figure}

{\textit{Circularity} -- The aggregates used in the experiments are far from being spherical. The deviation of the aggregate 
shape from the spherical shape is expressed by the circularity $c$ which,} is defined as
\begin{equation}
 c = \frac{A_{\rm agg}}{A_{\rm max}}
 \label{eq:circularity}
\end{equation}
where $A_{\rm agg}$ is the projected aggregate area, $A_{\rm max} = \pi \, r^2_{\rm max}$ is the area of a circle and, $r_{\rm max}$ is the maximum distance between the center of mass and the edges of the aggregate ({Fig.} \ref{fig:circularity}).

\begin{figure}
\resizebox{\hsize}{!}{\includegraphics{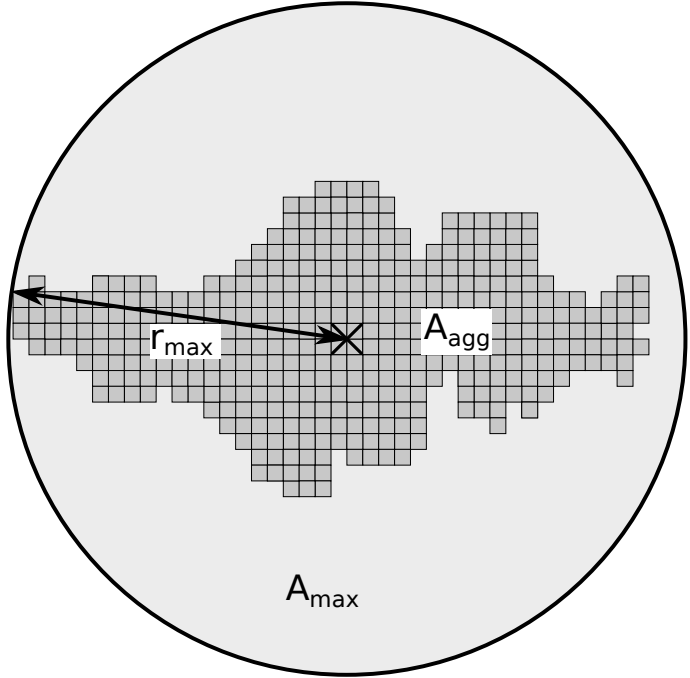}}
\caption{The circularity $c$ of an aggregate is defined by the relation of the aggregates area $A_{\rm agg}$ (dark-gray) and the maximal area $A_{\rm max}$ (light-gray) of a circle defined through $r_{\rm max}$, which is the maximum distance between an aggregate pixel and the center of mass (black cross).}
\label{fig:circularity}
\end{figure}

{\textit{Impact velocity} -- } The impact velocity $v$ is calculated from the distances between the aggregate centers of mass. The centers of mass of the aggregates are calculated directly from the discrete pixels of the projected aggregate areas via $r_{\rm com} = \sum_i r_i \cdot n^{-1}$, while $r_i$ are the respective coordinates of the aggregate pixels and $n$ is the total number of aggregate pixels. Owing to the gas outflow on the sides below the aggregates, the approach of two aggregates is slightly decelerated. We used a second order polynomial to fit the approach of the aggregates. An example of this fit can be seen in Fig. \ref{fig:approach}.

\begin{figure}
\resizebox{\hsize}{!}{\includegraphics{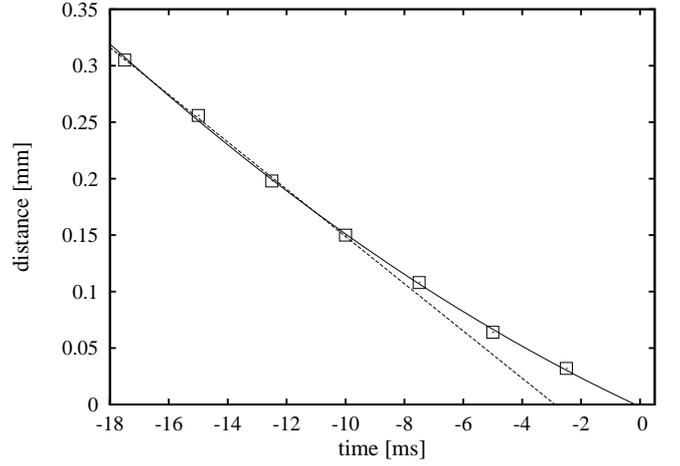}}
\caption{Example approach of two colliding aggregates with a collision velocity at $t = 0 \, \rm ms$ of $v \approx 1.8 \, \rm cm\,s^{-1}$. The movement is approximated by a second order polynomial (solid line). In comparison, a linear fit to the first four data points is shown to visualize the decelerating component of the sideflow of gas, which is induced by the Knudsen effect.}
\label{fig:approach}
\end{figure}

The collision velocity depends on -- besides the accuracy of the fitting parameters -- the exact point in time when the two aggregates touch. The collision time is defined through the contact of the two aggregates within the 2D image sequence with an uncertainty of one image or 2.5 ms. For the slowest collisions (1 cm/s) with a low total mass, this implies a maximum uncertainty of 6.5~\%. For fast collisions with high total masses, the deceleration is less intense and the velocities are more accurate with uncertainties of up to 3 \%.

{\textit{Maximum contact area }-- } The contact area is important in particle collisions as it determines the number of 
sticking connections and the way in which the force is distributed among the grains. We consider the maximum 
contact area $A_{\rm c}$ of the aggregates visible in the 2D images. If the maximum contact line $l_{\rm c}$ (see Fig. \ref{fig:contactarea}) during a collision were equivalent to the diameter of a circle, the circular contact area would be $\pi \, l_{\rm c}^2 \cdot 1/4$. The aggregate's bottom side is flat because it was lifted from a plane surface. Hence, the assumption of a full circle as a contact area leads to an overestimate of the contact area by a factor of two giving $\pi \, l_{\rm c}^2 \cdot 1/8$. As the lateral extension of the aggregates is usually larger than the height, we assume a more ellipsoid shape of the contact area, which adds a small factor of 0.8 to the maximum contact area, a conservative improvement but otherwise arbitrary. Finally, the maximum contact area is
\begin{equation}
 A_{\rm c} = \frac{\pi \, l_{\rm c}^{2}}{10}.
 \label{eq:contactareacalc}
\end{equation}

\begin{figure}
\resizebox{\hsize}{!}{\includegraphics{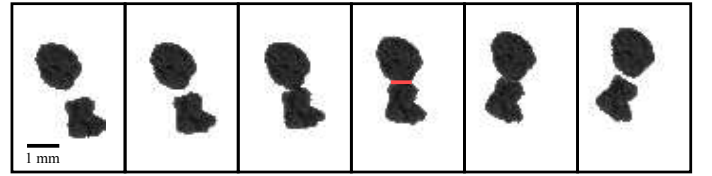}}
\caption{Example of a collision between two aggregates with a time step of $\sim 8$ ms from the left to the right. The red (gray) line corresponds to the maximum contact line $l_{\rm c}$.}
\label{fig:contactarea}
\end{figure}

\textit{Impact parameter } -- We define an impact parameter $I$ to describe the 
difference between central and oblique collisions as
\begin{equation}
 I = \frac{d}{r_{\rm 1} + r_{\rm 2}}.
 \label{eq:impactparameter}
\end{equation}
This is a dimensionless value between 0 (central collision) and 1 (contact only on the outer edges of both aggregates), where $d$ is the distance between the projected trajectory of the smaller aggregate and the center of mass of the larger aggregate (Fig. \ref{fig:impactparameter}). The dimensions $r_{\rm 1}$ and $r_{\rm 2}$ are defined to be perpendicular to the impact direction one frame before the aggregates come into contact.

\begin{figure}
\resizebox{\hsize}{!}{\includegraphics{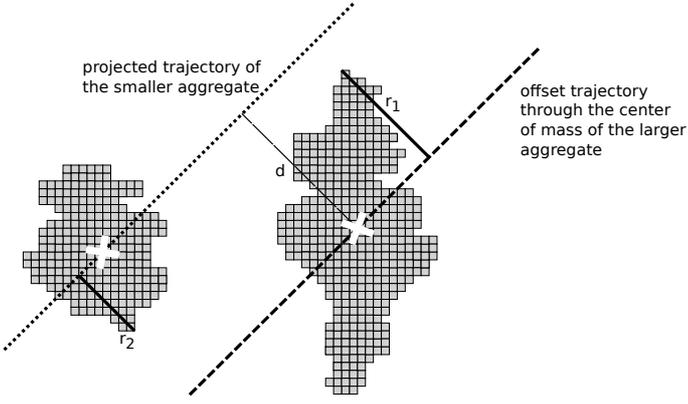}}
\caption{Visualization of the components used for the impact parameter calculation, where $r_1$, $r_2$ and $d$ are determined on the last frame before the aggregates get in touch. The dotted line represents the momentary projected trajectory of the smaller aggregate relative to the center of mass of the larger aggregate (white cross). The offset trajectory through the center of mass of the larger aggregate is marked with a dashed line. The distance $d$ between the center of mass of the larger aggregate and the momentary projected trajectory of the smaller aggregate is visualized with a thin line.}
\label{fig:impactparameter}
\end{figure}

{\textit{Impact energy} -- } The impact energy $E$ is calculated using the impact velocity {$v$ and the calculated masses $m_1$ and $m_2$} via
\begin{equation}
E = \frac{1}{2}\frac{m_1m_2}{m_1 + m_2}v^2.
 \label{eq:impactenergy}
\end{equation}

\section{Results and discussion}
After the data reduction, the following parameters of a collision are given: the aggregate masses $m_1$ and $m_2$, the circularities $c_1$ and $c_2$ of the aggregates, the impact velocity $v$, the impact energy $E$, the maximum contact area $A_{\rm c}$, and the impact parameter $I$. In the following, we divide the collisions into two types:
\begin{description}
\item \textit{Normal collisions} -- These are collisions between the aggregates that were placed initially on the heater (a total of 239).
\item \textit{Pre-accreted collisions} -- These are collisions between aggregates that
were formed before by a sticking collision (a total of 101).
\end{description}
The latter collisions are clearly visible as two connected aggregates. Most basic models of dust aggregate collisions in a protoplanetary disk provide a collision velocity depending on the two particle masses (and sizes). In Fig. \ref{fig:parspacenormal}, the outcomes of the individual normal collisions and in Fig. \ref{fig:parspaceacc} those of the pre-accreted aggregates are plotted depending on the collision velocity $v$ and the sum of the mass of the two aggregates (green: sticking; yellow: bouncing; red: detachment).\\

\begin{figure}
\resizebox{\hsize}{!}{\includegraphics{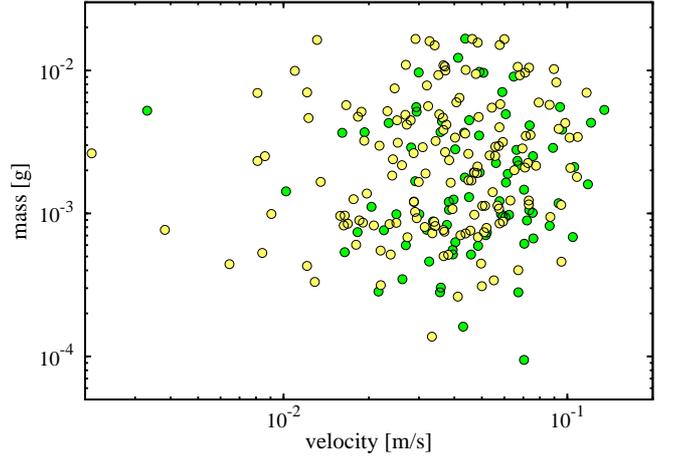}}
\caption{Outcome of the individual normal collisions: green indicates sticking particles and yellow those bouncing.}
\label{fig:parspacenormal}
\end{figure}

\begin{figure}
\resizebox{\hsize}{!}{\includegraphics{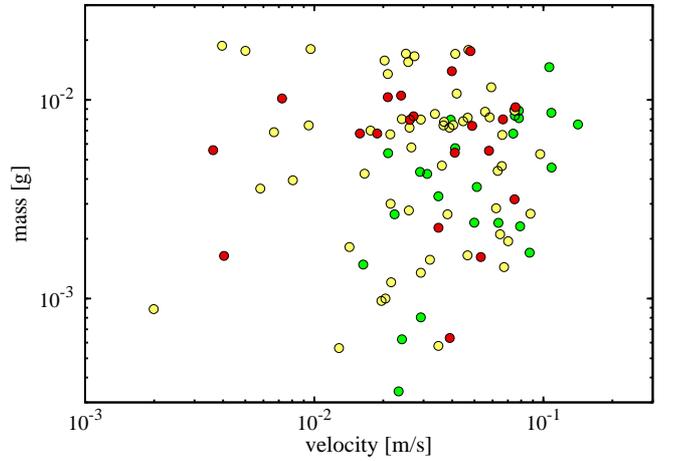}}
\caption{Outcome of the individual pre-accreted collisions: green sticking particles, yellow those bouncing, and red detachment.}
\label{fig:parspaceacc}
\end{figure}

The sum of the aggregate masses range between $10^{-1}\, {\rm mg} $ and $20\, {\rm mg}$ and their velocities between $0.3 \, {\rm cm/s}$ and $15 \, {\rm cm/s}$. For all 340 collisions, sticking occurs with a probability of $32.1 \pm 2.5$~\%. The sticking probability for normal collisions is on average $35.1 \pm 3.0$~\% and for pre-accreted collisions $24.8 \pm 5.3$~\%. The probability that a collision between two pre-accreted aggregates leads to detachment is $19.8 \pm 4.0$~\%. In the following, we evaluate how the sticking probability depends on the different parameters mentioned above. In general, we give three probabilities, one probability for sticking $p_{\rm s}$, one for bouncing $p_{\rm b}$ and one for detachment $p_{\rm d}$. The number of data points in a bin is typically 25, depending slightly on the chosen data set. The error bars in the probabilities mark the standard deviation in the mean. Error bars of certain quantities also include measurement errors. At the original spatial resolution of $\sim 50$~$\mu$m/pixel, we do not observe splits of dust aggregates in collisions but at high spatial resolution ($\sim 7.5$~$\mu$m/pixel) we see small fragments or small-scale mass transfer from one aggregate to another (Fig. \ref{fig:highres}). Partial compaction of aggregates was also observed in the high-resolution configuration.

\begin{figure}
\resizebox{\hsize}{!}{\includegraphics{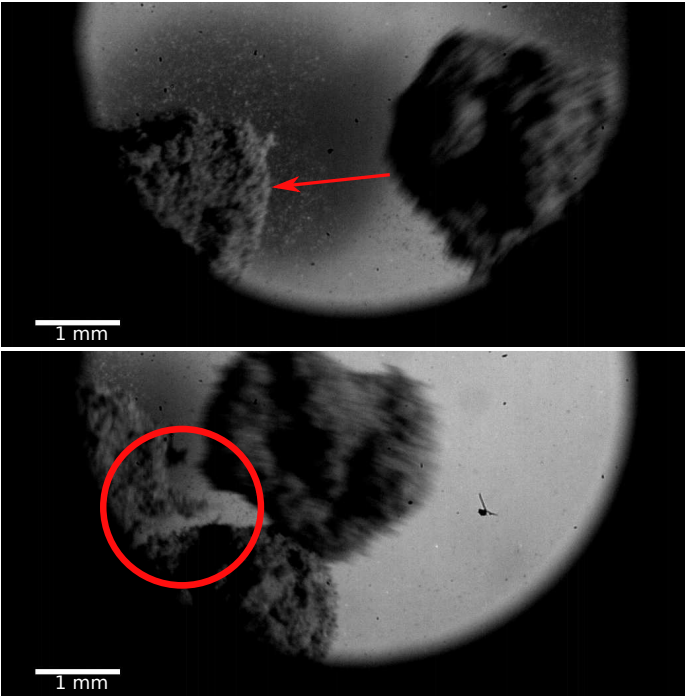}}
\caption{High spatial-resolution record of a collision with a mass transfer at $\approx 5.2$ cm/s. A small fragment from the right aggregate is attached to the left aggregate (circle) after the collision.}
\label{fig:highres}
\end{figure}

The velocity dependence of the sticking probability of the 239 normal and the 101 pre-accreted aggregate collisions are depicted in Figs.\ref{fig:probvelocity} and \ref{fig:probvelocity2}. There is a clear tendency for the sticking probability to increase with collision velocity. We used the following analytic expression to fit the data for normal collisions
\begin{equation}
 p(v) = \alpha_{\rm v} \, \left(1 - {\rm e}^{\,-\beta_{\rm v} \, v} \right),
\label{eq:fitfunctionexp}
\end{equation}
with $\alpha_{\rm v} = 0.47 \pm 0.07$ and $\beta_{\rm v} = (0.32 \pm 0.10) \,\rm s \, cm^{-1}$.
The reason behind this choice is as follows: for small velocities, we expect bouncing as no restructuring occurs and the collisions are elastic. For large velocities, the sticking probability has to level off at a value smaller than one by definition. Equation (\ref{eq:fitfunctionexp}) is a simple function that fulfills these requirements. For pre-accreted aggregates, detachment also occurs. At higher velocity, the detachment probability is significant lower however than the sticking probability.

\begin{figure}
\resizebox{\hsize}{!}{\includegraphics{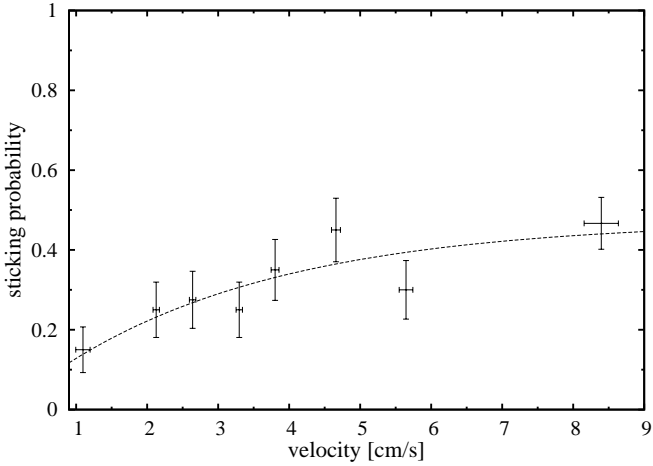}}
\caption{Velocity dependence of the sticking probability of normal collisions. The dotted line corresponds to $p_{\rm s}(v) = 0.47 \cdot (1 - e^{-0.32\,v})$.}
\label{fig:probvelocity}
\end{figure}

\begin{figure}
\resizebox{\hsize}{!}{\includegraphics{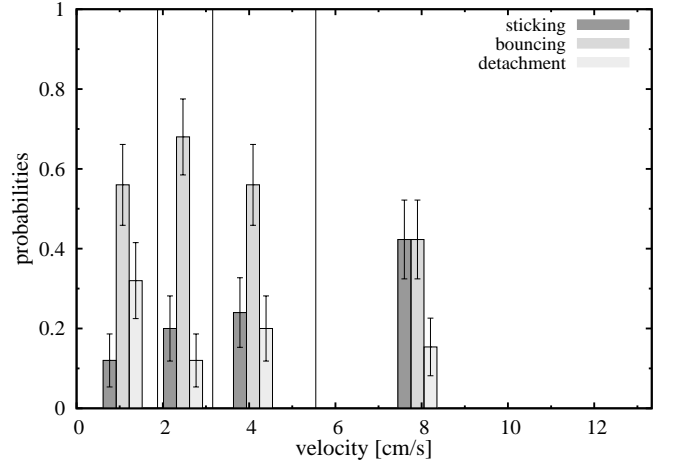}}
\caption{Velocity dependence of the sticking probability of pre-accreted collisions (dark-gray: accretion; gray: bouncing; light-gray: detachment). The colored bars represent the probabilities and were placed at the average velocity of the respective bins, which are indicated by the vertical lines.}
\label{fig:probvelocity2}
\end{figure}

The overall (normal and pre-accreted) sticking probability dependence on the collision energy is shown in Fig. \ref{fig:probenergy}. There is only a slight trend that higher energy leads to an increase in the sticking probability, but for irregular particles the contact areas vary and play an important role in the energy distribution and dissipation. We therefore define an energy density as $ E_{\rm d} = E / A_{\rm c}$, where $ E $ is the impact energy and $ A_{\rm c} $ is the maximum contact area. The larger his value, the more the restructuring that can occur to dissipate the energy that might be beneficial for sticking. Figure \ref{fig:probenergydensity} indeed shows a clear increase \textbf{in} the sticking probability (normal and pre-accreted) with energy density in the investigated regime. We used the same function (Eq. \ref{eq:fitfunctionexp2}) to fit the data
\begin{equation}
 p(E_{\rm d}) = \alpha_{\rm E_{d}} \, \left(1 - {\rm e}^{\,-\beta_{\rm E_{d}} \, E_{\rm d}} \right)
\label{eq:fitfunctionexp2}
\end{equation}
with $\alpha_{\rm E_{d}} = 0.40 \pm 0.02$ and $\beta_{\rm E_{d}} = (13.4 \pm 2.6) \, \rm mm^{2} \, nJ^{-1}$.

\begin{figure}
\resizebox{\hsize}{!}{\includegraphics{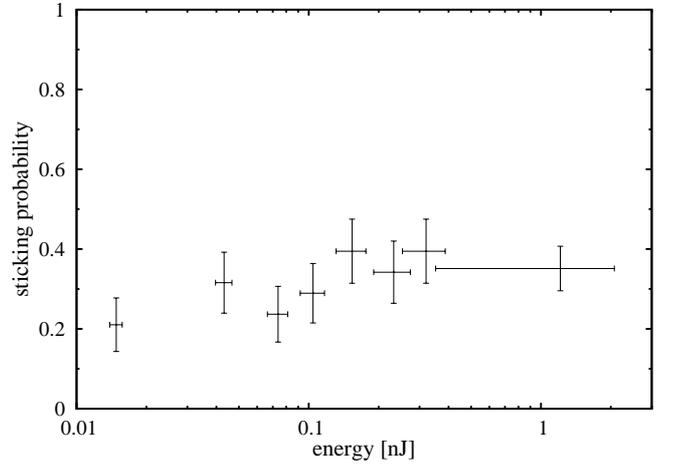}}
\caption{Impact energy dependence of the sticking probability for normal and pre-accreted collisions.}
\label{fig:probenergy}
\end{figure}

\begin{figure}
\resizebox{\hsize}{!}{\includegraphics{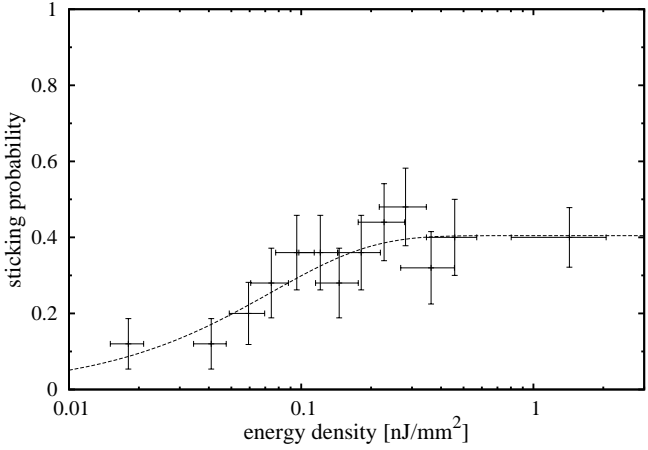}}
\caption{Impact energy density dependence of the sticking probability $E_{\rm d}$ of all collisions, where the appropriate function is $p_{\rm s}(E_{\rm d}) = 0.40 \cdot (1 - e^{-13.4 \, E_{\rm d}})$.}
\label{fig:probenergydensity}
\end{figure}

Collisions can also be characterized by the mass ratio of the impacting particles. The sticking probability for normal collisions is depicted in Fig. \ref{fig:probmassratio}. Within the studied interval from equal masses to a ratio of 100, a significant dependence is not visible.

\begin{figure}
\resizebox{\hsize}{!}{\includegraphics{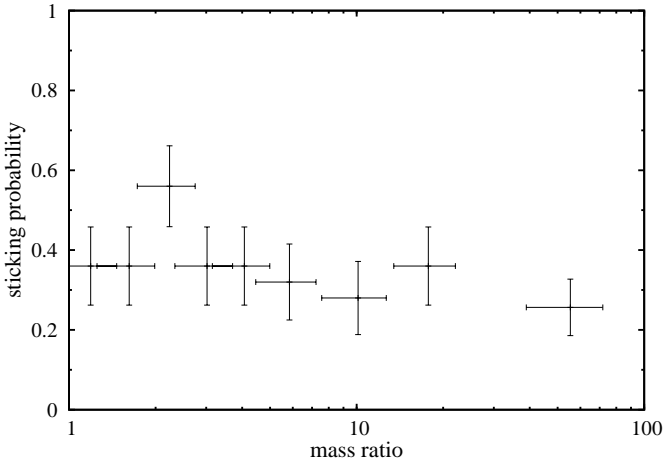}}
\caption{Sticking probability versus mass ratio for normal collisions.}
\label{fig:probmassratio}
\end{figure}

To see a potential shape effect, we show the sticking efficiency on the average circularity $c_{avg}=(c_1+c_2)/2$ for normal collisions in Fig. \ref{fig:probcircularity}. There is no trend in the data but instead a rather large variation.

\begin{figure}
\resizebox{\hsize}{!}{\includegraphics{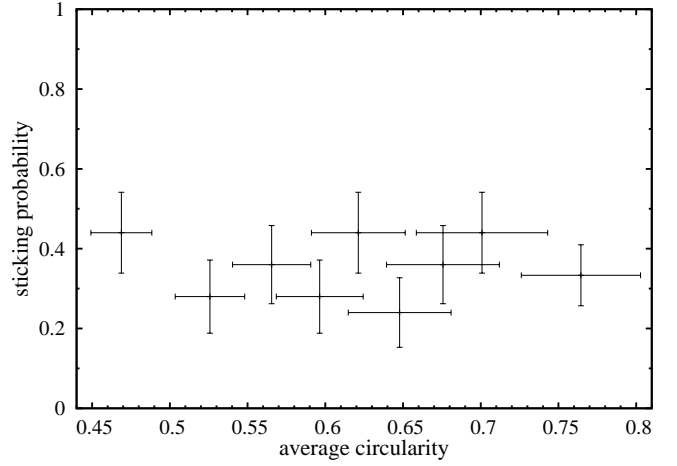}}
\caption{Dependence of the sticking probability of normal collisions on the average circularity $c_{avg}$.}
\label{fig:probcircularity}
\end{figure}

The last dependence considered is on the impact parameter $I$, which in general is known to be important in collisions. The probabilities are plotted versus the impact parameter for both normal collisions in Fig. \ref{fig:probimpactparameter1} and pre-aggregated collisions in \mbox{Fig. \ref{fig:probimpactparameter2}}, where a clear dependence is visible. For normal collisions (Fig. \ref{fig:probimpactparameter1}), the sticking probability first decreases to a minimum at $I \approx 0.2$. This could be because head-on collisions can easily lead to deformation but with increasing impact parameter these deformations were diminished. The sticking probability then increases to a maximum at $I \approx 0.3$. We attribute this to the possibility of dissipating energy by inducing particle rotation. The sticking probability then decreases toward larger impact parameters because the normal velocity component gets smaller with less energy dissipation, while the tangential force at the same time is eventually sufficient to allow sliding with some friction but inertia to dominate. A simple analytic expression to quantify the decreasing part is
\begin{equation}
 p_{\rm s}(I) = - \gamma + \frac{1}{I + \tau}
\label{eq:fitfunctiontwo}
\end{equation}
with $\gamma = 0.38 \pm 0.02$ and $\tau = 0.84 \pm 0.04$.

For pre-accreted collisions, the behavior is quite similar (see Fig. \ref{fig:probimpactparameter2}). First the sticking probability increases to a maximum at $I \approx 0.3$ and then decreases toward larger impact parameters. The detachment probabilities for pre-accreted collisions seem to be slightly larger at larger impact parameters. This would be in agreement with inducing rotation and tangential forces between sticking aggregates that would more easily break the contact. The effect is not strong and an increase is statistically not observed at a significant level.

\begin{figure}
\resizebox{\hsize}{!}{\includegraphics{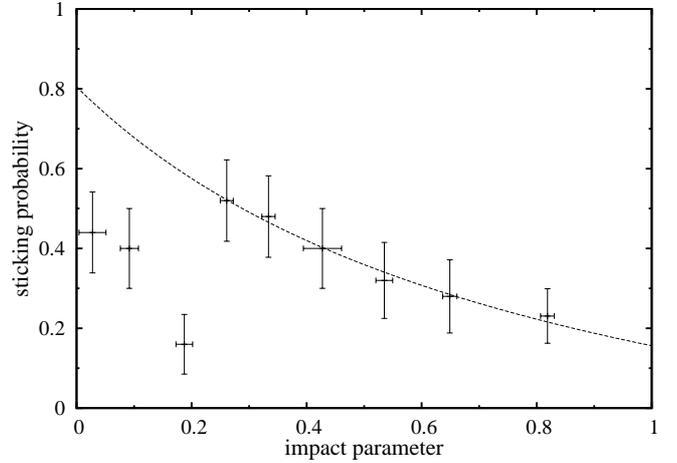}}
\caption{Sticking probability versus impact parameter for normal collisions. The dotted line is a fit for $I > 0.2$ with $p_{\rm s}(I) = - 0.38 + (I + 0.83)^{-1}$.}
\label{fig:probimpactparameter1}
\end{figure}

\begin{figure}
\resizebox{\hsize}{!}{\includegraphics{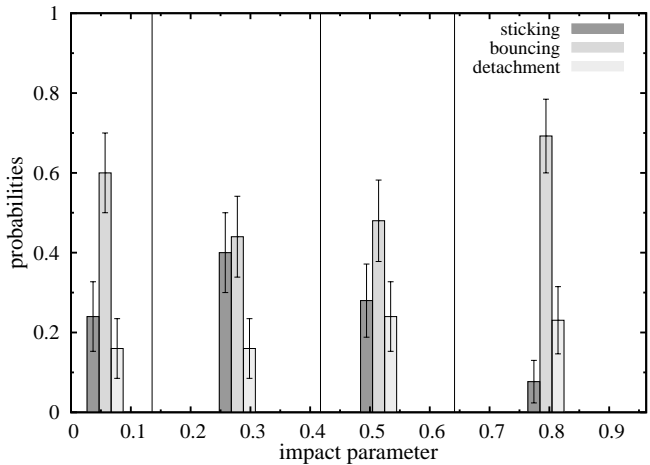}}
\caption{Sticking probability \textbf{versus} impact parameter for pre-accreted aggregates (dark-gray: accretion; gray: bouncing; light-gray: detachment).}
\label{fig:probimpactparameter2}
\end{figure}

At very low velocities, repulsive gas drag might lead to a sticking probability that is systematically lower. In addition a number of detachments are likely supported by the repulsive gas drag as we also observed detachment without any further collision.  We did not record a side view with every collision and aggregate and at the highest impact energies in particular aggregates might slide over each other, which would lead to sticking where free collisions would not necessarily result in sticking. Collisions for which this bias was obvious in the videos were not taken into account here but we cannot exclude these effects on a non-resolvable scale. This would lead to a systematic increase in the sticking efficiency. On the other hand, that the sticking efficiency depends on the impact parameter also suggests that the sticking enhancements due to the two-dimensional setup do not dominate our results as other dependences should not then be so pronounced and agree with general expectations.

To summarize our results, our experiments show that collisional outcomes (sticking probabilities) in the studied parameter range depend significantly on only three parameters: impact parameter, collision energy density, and collision velocity.

\section{Conclusion}
There is currently a debate about whether the growth of dust would stall at mm-size, which has been termed the bouncing barrier by \citet{zsom2010}. This barrier would arise if particles entered a regime of equally sized mm particles colliding only among themselves at velocities $\lesssim$ 1 m/s, where only bouncing was the result of a collision. The idea is already present in early laboratory experiments \citep{blum1993}. Rebound is also seen in n-particle simulations depending on the coordination number \citep{wada2011}. Numerical modeling, including all different kinds of collisions then result in a bouncing barrier \citep{zsom2010}. However, the most pronounced input is that only rebound was observed in collisions of mm-size aggregates in the relevant domain of low collision velocities \citep{heisselmann2010, weidling2011}. The experiments that are important used dust with grain sizes of about $1 \mu {\rm m}$. This is reasonable as for instance matrix material in many pristine meteorites contain particles of this size \citep{brearley1999}. It is also reasonable as dust aggregates 
of this size are subject to strong cohesive forces that might be beneficial to growth by means of sticking. However, for compact mm-aggregates the strong sticking might be counterproductive for further growth. Aggregates are so strong in slow collisions that no energy dissipation by restructuring can occur. The collisions then get mostly elastic and rebound is the outcome.

That rearrangement of grains is beneficial to growth has e.g. been shown by impacts of cm-size aggregates at velocities of up to 2~m/s \citep{beitz2011}, 1.5 - 6 m/s \citep{kothe2010}, or even 60 m/s  \citep{teiser2009b}. These experiments show that as soon as the destruction of a projectile is possible, some mass might be transferred to a somewhat larger target, leading to net growth.

Slow collision experiments with larger individual grains were carried out by \citet{hartmann1978} and \citet{colwell2008} showing that larger solid projectiles might stick to the dust target in the sub m/s range. Obviously going to a more granular medium with less cohesive forces might help growth at least at certain stages. In addition, the recent experiments of \citet{beitz2011b} on mm-size particles with dust rims assume that including larger particles helps us to dissipate energy and increase sticking velocities. In principle, varying the monomer sizes shifts the velocity ranges where certain collisional regimes such as sticking, bouncing, and fragmentation dominate.

The grains used in our experiments are up to $25~\mu{\rm m}$ in size. With a $54$\% $\pm$ $1$\% porosity, the aggregates are already rather dense and unlikely to be compacted much further in low speed collisions. Though aggregates consisting of these particles are fragile, they still experience sufficient sticking to form aggregates. What they provide by their low sticking forces is a mean energy dissipation in the slow collisions that leads to the large observed sticking probabilities. Therefore, if present, larger grains will aid collisional growth at low velocities, which might bridge growth to particles of cm-size or larger.

\textit{Caveats} -- (1) If the experimental results are to be applied to planetesimal formation, particle size is very important. It is currently still an unknown property of particles in protoplanetary disks but the basalt particles used are likely among the largest of possible grain sizes (excluding chondrules). Nevertheless, if only a few aggregates consisting of large grains formed in the disk just by statistical reasoning, they might have acted as seeds for further growth but this has to be studied in further experiments.

(2) More fundamentally, one has to be aware that our experiments were confined to two dimensions and the aggregates were supported against gravity. However, the observed trends agree with expectations, e.g. the dependence on the 2D impact parameter would also be expected in a three-dimensional case. This indicates that the high sticking probabilities are no artifacts of the experiment in general.

We were unable to determine whether accretion predominates over detachment based on the experiment so far and the bouncing barrier might still exist for most of the collisions. Nevertheless, 
sticking can be efficient and allows the formation of larger aggregates. By simple statistical reasoning, as not every aggregate becomes detached by the subsequent collisions, the growth of larger particles seams feasible. The model of \citet{windmark2012} requires some seeds but not too many, a situation that the collisions of aggregates with larger constituents might help to improve.

\section{Acknowledgements}

This work is funded by the DFG research group FOR 759.  

\bibliographystyle{aa} 
\bibliography{2012_aanda.bib} 

\end{document}